\documentclass[preprint, double-spaced]{revtex4}
\usepackage{graphicx}
\usepackage{amsmath,bbm}
\usepackage{amssymb,bm}

\begin{document}

\title{Baryon-Size Dependent Location of QCD Critical Point}
\author{P.~K.~Srivastava}
\author{S.~K.~Tiwari}
\author{C.~P.~Singh\footnote{corresponding author: $cpsingh_{-}bhu@yahoo.co.in$}}

\affiliation{Department of Physics, Banaras Hindu University, 
Varanasi 221005, INDIA}

\begin{abstract}
\noindent
The physics regarding the existence of the critical end point (CEP) on the QCD phase boundary still remains unclear and its precise location is quite uncertain. In this paper we propose that the hard-core size of the baryons used in the description of the hot and dense hadron gas (HG) plays a decisive role in the existence of CEP. Here we construct a deconfining phase transition using Gibbs' equilibrium conditions after using a quasiparticle equation of state (EOS) for QCD plasma and excluded-volume EOS for the HG. We find that the first order transition results only when we assign a hard-core size to each baryon in the description of HG and the phase boundary thus obtained terminates at CEP beyond which a cross-over region occurs. The mean field approach for the HG lends support to this finding where unless we include an excluded-volume effect in the approach, CEP does not materialize on the QCD boundary. This investigation provides an intuitive reasoning regarding the origin of CEP and the cross-over transition on the QCD phase boundary.
\\


\end{abstract}
\maketitle 
\section{ Introduction}

\noindent
Precise mapping of the QCD phase boundary existing between two distinct phases of hot, dense hadron gas (HG) and weakly interacting plasma of quarks and gluons (QGP) and the location of hypothesized critical end point (CEP) have emerged as interesting and challenging problems before the experimental and theoretical heavy-ion physicists today [1-3]. The discovery of QCD critical end point is bound to clear the mist surrounding our understanding of the conjectured QCD phase diagram and hence it would help us to ascertain the properties and signals of QGP [4-6] to some extent. The possible existence of CEP in the temperature ($T$) and baryon chemical potential ($\mu_{B}$)plane of the QCD phase boundary was proposed a decade ago and it represents a second-order transition point where the first-order transition boundary terminates as $T$ increases and $\mu_{B}$ decreases [3]. Its separation from the temperature axis $(\mu_{B}=0)$ spans the region of a cross-over transition . Lattice QCD studies employing Monte Carlo simulation have failed at $\mu_{B}\ne0$ because the absence of a probability measure precludes direct computations and hence many mathematical approximations have been used to determine QCD phase diagram for nonvanishing values of $\mu_{B}$. Some of these calculations predict the CEP to occur in the range $\mu_{C}/T_{C}\approx 1.0-1.4$ [7-8]. However, certain calculations have also cast a shadow of doubt over the occurrence of a CEP in the phase diagram [9]. We still do not have any intuitive picture for understanding the circumstances under which a cross-over transition can occur around $\mu_{B}=0$ and which finally culminates into a CEP as $\mu_{B}$ increases. This is qualitatively supported by some lattice QCD findings [7-8]. In this paper, we take the help of a phenomenological model to emphasize the dominant role played by the finite-size baryons as constituents of a hot, dense HG in the existence of a cross-over as well as CEP on the QCD phase boundary. \\
The search for a realistic equation of state (EOS) of HG is essential for the proper understanding of the properties of the QGP. In a simple treatment of HG, all the baryons and mesons are treated as non-interacting point-like particles. However, such EOS of the HG has an undesirable feature that at very large $\mu_{B}$ and/or $T$, the hadronic phase reappears as a stable configuration in the Gibbs' construction of equilibrium phase transition between HG and QGP. Thus the pressure equality condition $P_{HG}=P_{QGP}$ occurs twice for two values of $\mu_{B}$ for each $T$ because of an exponential growth of hadrons and their resonances at higher $\mu_{B}$ [10]. Thus an anomalous feature of the reversal of phase transition from QGP to HG occurs which is indeed contrary to what we expect from the asymptotic freedom of QCD. In literature, this feature was handled by modelling the repulsive interactions existing between hadrons at large $T$ and/or $\mu_{B}$ of HG either in the mean field approach or in the excluded volume method. The attractive and repulsive interactions in the mean fields are incorporated in HG by scalar $\sigma$ and vector $\omega$-exchange, respectively [11]. The Yukawa potential due to $\omega$-exchange $V(r)=(G^{2}/4\pi r)exp(-m_{\omega}r)$ generates a mean potential energy in HG as $U_{B}=G^{2}n_{B}/m_{\omega}$ which vanishes when the net baryon density $n_{B}\rightarrow 0$. This means that one can again generate a large number of hadronic resonances at large $T$ where $n_{B}=0$ and consequently $P_{HG}$ again becomes larger than $P_{QGP}$ at very large $T$. In the recent past, we have attempted to cure this problem by adding a Vander-Waals repulsive interaction term $U_{VDW}(n,T)$ which depends on $T$ and $n$ (total number density of particles) and has its origin in the excluded volume correction [12, 13]. However, the main drawback of this model is that we cannot mathematically handle the EOS of HG if it incorporates many resonances in its description and a large uncertainty also results due to many unknown coupling parameters entering into the EOS. An alternate phenomenological description has mostly been used in the description of HG and the repulsive force arising due to hard-core volume of each baryon is geometrically incorporated as excluded volume correction in the pressure of HG [10, 14-16]. After such correction, we obtain a reduced pressure and hence $P_{HG}=P_{QGP}$ is satisfied for only one value of $\mu_{B}$ at each $T$. In order to obtain $P_{QGP}$, we either use a simplistic bag model EOS or we take the help of more realistic quasiparticle model. Recently we pointed out that QCD phase boundary obtained from such exercise depicts a first order deconfining phase transition and the transition line indeed terminates at CEP beyond which a cross-over transition occurs when $T$ further increases and $\mu_{B}$ decreases slightly [17-18].\\
In this paper, we plan to investigate the important role of baryon size as the origin in the existence of CEP and the resulting cross-over transition. We seek an answer to the question: why should a CEP occur on the conjectured QCD boundary and to what extent, the modelling of the HG should be held responsible for the existence of CEP?\\
The rest of the paper is organised as follows. In section II, we discuss briefly the main features of the excluded volume models used in this paper. We also give outlines of our thermodynamically consistent formulation of the excluded volume model which was also used in the previous papers [17, 18]. Section III deals with the mean field approach used for the description of HG where we incorporate an explicit term depicting the excluded volume correction. In section IV, we give a quasiparticle model which is a thermodynamically consistent formulation of the EOS for QGP. Finally, section V gives our results, detailed comparisons and conclusions.
 
\section{Formulation of excluded volume model for HG}
\noindent
In the excluded volume models, repulsive interaction between two hadrons has been included by giving the hadrons a hard-core geometrical size and consequently reduces the hadronic degrees of freedom at large $T$ and/or $\mu_{B}$. Consequently, the hadronic pressure is reduced and one can get a deconfinement phase transition from HG to QGP by using Gibbs' construction. However, some of these models are not thermodynamically consistent because number density cannot be obtained directly from the partition function. There were many attempts to obtain a thermodynamically consistent formulation of the excluded volume model [10, 14-16]. Recently we have proposed a thermodynamically consistent excluded volume model for hot and dense HG [17, 18]. Our approach has the following new features. Besides thermodynamical consistency, our model uses full quantum statistics so that the phase boundary in the entire $(T, \mu_{B})$ plane can be obtained without using any additional approximation. Recently, we have used our model for constructing a first order deconfining phase boundary and the phase boundary is found to terminate at CEP [17, 18]. We have further determined the chemical freeze-out curve from our HG model and its proximity to CEP was pointed out [1]. In this model, we give importance to baryonic hard-core repulsion and we thus incorporate excluded-volume correction arising due to baryonic size only. We assume that mesons can overlap and fuse into one another and hence do not possess any hard-core repulsion. The grand canonical partition function for the HG, with full quantum statistics and after incorporating excluded volume correction in this model can be explicitly written as:
\begin{equation}
\begin{split}
ln Z_i^{ex} = \frac{g_i}{6 \pi^2 T}\int_{V_i^0}^{V-\sum_{j} N_j V_j^0} dV
\\
\int_0^\infty \frac{k^4 dk}{\sqrt{k^2+m_i^2}} \frac{1}{[exp\left(\frac{E_i - \mu_i}{T}\right)+1]}
\end{split}
\end{equation}
where $g_i$ is the degeneracy factor of ith species of baryons, $E_{i}$ is the energy of the particle ($E_{i}=\sqrt{k^2+m_i^2}$), $V_i^0$ is the eigenvolume of one baryon of ith species and $\sum_{j}N_jV_j^0$ is the total occupied volume by the baryons and $N_{j}$ represents total number of baryons of jth species.

Now we can write Eq.(1) as:

\begin{equation}
ln Z_i^{ex} = V(1-\sum_jn_j^{ex}V_j^0)I_{i}\lambda_{i},
\end{equation}
where $I_{i}$ represents the integral:
\begin{equation}
I_i=\frac{g_i}{6\pi^2 T}\int_0^\infty \frac{k^4 dk}{\sqrt{k^2+m_i^2}} \frac1{\left[exp(\frac{E_i}{T})+\lambda_i\right]},
\end{equation}
and $\lambda_i = exp(\frac{\mu_i}{T})$ is the fugacity of the particle, $n_j^{ex}$ is the number density of jth type of baryons after excluded volume correction and can be obtained from Eq.(2) as:
\begin{equation}
n_i^{ex} = \frac{\lambda_i}{V}\left(\frac{\partial{ln Z_i^{ex}}}{\partial{\lambda_i}}\right)_{T,V}
\end{equation}
This leads to a transcendental equation as
\begin{equation}
n_i^{ex} = (1-R)I_i\lambda_i-I_i\lambda_i^2\frac{\partial{R}}{\partial{\lambda_i}}+\lambda_i^2(1-R)I_i^{'}
\end{equation}
where $I_{i}^{'}$ is the partial derivative of $I_{i}$ with respect to $\lambda_{i}$ and $R=\sum_in_i^{ex}V_i^0$ is the fractional occupied volume. We can write R in an operator equation as follows [10]:
\begin{equation}
R=R_{1}+\hat{\Omega} R
\end{equation}
where $R_{1}=\frac{R^0}{1+R^0}$ with $R^0 = \sum n_i^0V_i^0 + \sum I_i^{'}V_i^0\lambda_i^2$; $n_i^0$ is the density of pointlike baryons of ith species and the operator $\hat{\Omega}$ has the form :
\begin{equation}
\hat{\Omega} = -\frac{1}{1+R^0}\sum_i n_i^0V_i^0\lambda_i\frac{\partial}{\partial{\lambda_i}}
\end{equation}
Using Neumann iteration method and retaining the series upto $\hat{\Omega}^2$ term, we get
\begin{equation}
R=R_{1}+\hat{\Omega}R_{1} +\hat{\Omega}^{2}R_{1}
\end{equation}
\noindent
Eq.(8) can be solved numerically for R. Finally, we get the total pressure [17, 18] of the hadron gas:
\begin{equation}
\it{p}_{HG}^{ex} = T(1-R)\sum_iI_i\lambda_i + \sum_j\it{p}_j^{meson}
\end{equation}

In (9), the first term in the right hand side represents the pressure due to all types of baryons where excluded volume correction is incorporated and the second term gives the total pressure due to all mesons in HG having a pointlike size. In this calculation, we have taken an equal volume $V^{0}=\frac{4 \pi r^3}{3}$ for each type of baryon with a hard-core radius $r=0.8 fm$. We have taken all baryons and mesons and their resonances having masses upto $2 GeV/c^{2}$ in our calculation for HG pressure. We have also used the condition of strangeness neutrality by putting $\sum_{i}S_{i}(n_{i}^{s}-\bar{n}_{i}^{s})=0$, where $S_{i}$ is the strangeness quantum number of the ith hadron, and $n_{i}^{s}(\bar{n}_{i}^{s})$ is the strange (anti-strange) hadron density, respectively. We want to stress here that the form of this model used under Boltzmann approximation has been found to describe [21] the observed multiplicities and the ratios of the particles in heavy-ion collisions. In order to show the comparison of our results with the results obtained in a thermodynamically inconsistent approach of Cleymans and Suhonen [19] which has been described in detail in the ref. [10]. In this approach, the excluded baryon density of ith species can be written as [20]:
\begin{equation}
n_{i}^{ex}=\frac{n_{i}^{0}}{1+\sum_{i}n_{i}^{0}V_{i}^{0}}
\end{equation} 
\section {Mean field Model for HG}
An alternate method for modelling the EOS for HG after incorporation of the repulsive interactions existing between hadrons of HG is the mean field approach. In this paper, we have used the mean field model of Tiwari et. al. [12], based on the work of Anchiskin and Suhonen [13], in which a Vander-Waals repulsive interaction term $U_{VDW}(n,T)$ is added. This term has its origin in the excluded volume correction. We extend this model to describe the interactions in the HG and include the contributions of the baryons $N , \Lambda, \sum, \Xi$ and $\Delta$-resonance in addition to the non-interacting mesons upto a cutoff mass of $2$ GeV in HG. We have again treated the mesons as pointlike particles. The attractive interaction between the baryons is given by the scalar $\sigma$-exchange while the exchange of vector $\omega$-meson gives the repulsive force. We have taken the value of coupling constant from one of the work of Suguhara and Toki [22] as follows
\begin{equation}
\frac{g_{\omega NN}}{m_{\omega}}=3.178 fm ,\;\; \frac{g_{\sigma NN}}{m_{\sigma}}=3.871 fm .
\end{equation}
We use SU(6) quark model to obtain the relations [23-28] between various couplings as
\begin{equation}
g_{\omega NN}=\frac{2}{3}g_{\omega\Lambda\Lambda}, \;\;g_{\omega NN}=\frac{2}{3}g_{\omega\sum\sum},\;\; \frac{1}{3}g_{\omega NN}=g_{\omega\Xi\Xi}.
\end{equation}
Moreover, the couplings for $\Delta$ particle are assumed to be the same as those of nucleons [29]. The expression for the total pressure of the HG in this mean field model, with excluded volume correction, can be written as [12, 13]
\begin{equation}
p=\frac{1}{3}\sum_{j}g_{j}\int \frac{d^{3}k}{(2\pi)^3}\frac{k^2}{(M_{j}^{*2}+k^{2})^{1/2}}\left[f_{j}+f_{\bar{j}}\right]+p_{VDW}(n,T)+p_{B}(n_{B})+p_{\sigma}(\sigma)+\sum_{m}p_{m}(T).
\end{equation}
In Eq. (13), the first term on the right hand side is the contribution from baryon with an effective mass $M_{j}^{*}$ and effective chemical potential $\mu_{j}^{*}$. Second term is the excess pressure because of the excluded-volume correction. Third term represents the baryon-density dependent mean-field pressure. Fourth term is the mean-field pressure due to $\sigma$-exchange and the last term on the right hand side is due to the contribution of pointlike mesons. Furthermore, the other terms involved in Eq. (13) are:
\begin{equation}
f_{j(\bar{j})}=\left[exp\left(\frac{(M_{j}^{*2}+k^{2})^{1/2}+U_{VDW}(n,T)\pm U_{Bj}(n_{B})\mp \mu_{j}}{T}\right)+1\right]^{-1}.
\end{equation}
Here the upper (lower) sign refers to baryons(anti-baryons), respectively. The expression for Vander-Waal hard-core repulsion terms $p_{VDW}$ and $U_{VDW}$ are:\begin{equation}
p_{VDW}(n,T)= nT\frac{V_{0}n}{1-V_{0}n},
\end{equation}
\begin{equation}
U_{VDW}(n,T)= T\frac{V_{0}n}{1-V_{0}n}-T ln(1-V_{0}n),
\end{equation}
with 
\begin{equation}
n= \sum_{j}(n_{j}+n_{\bar{j}}),
\end{equation}
and $V_{0}$ is the hard core volume of each baryon ($V_{0}=\frac{4}{3}\pi r_{0}^{3}$, $r_{0}=0.8 fm$). We have taken the same hard core volume for all type of baryons. Now the third term in the right hand side of Eq. (15) is represented as 
\begin{equation}
p_{B}=\frac{1}{2}m_{\omega}^{2}\omega_{0}^{2},
\end{equation}
where $\omega_{0}$ is the time component of the $\omega$-exchange field and in the mean field approximation, it is given by
\begin{equation}
\omega_{0}=\frac{1}{m_{\omega}^2}[g_{\omega NN}n_{BN}+g_{\omega\Lambda\Lambda}n_{B\Lambda}+g_{\omega\sum\sum}n_{B\sum}+g_{\omega\Xi\Xi}n_{B\Xi}+g_{\omega\Delta\Delta}n_{B\Delta}],
\end{equation}
where
\begin{equation}
n_{Bj}= n_{j}-n_{\bar{j}}=g_{j}\int\frac{d^{3}k}{(2\pi)^3}[f_{j}-f_{\bar{j}}],
\end{equation}
\begin{equation}
n_{B}=\sum_{j}n_{Bj}(\mu_{j},T).
\end{equation}
Here $g_{j}$ is the degeneracy factor. In order to calculate the net $\Delta$ number density, we use the relation [21]
\begin{equation}
n_{B\Delta}= n_{\Delta}-n_{\bar{\Delta}}=\frac{g_{\Delta}}{(2\pi)^3}\int_{0}^{\infty}W(M) dM\int d^{3}k[f_{\Delta}-f_{\bar{\Delta}}].
\end{equation}
Here $W(M)$ is the profile function which takes into account the finite width of $\Delta$-resonance [30]. Furthermore, we get $U_{Bj}$ in terms of the time component of vector field $\omega$ as
\begin{equation}
U_{Bj}(n_{B})=g_{\omega jj}\omega_{0}.
\end{equation}
The vector interaction of $\omega$-meson with all the baryons and $\Delta$-resonance modifies their chemical potential as
\begin{equation}
\mu_{j}^{*}=\mu_{j}-U_{Bj}.
\end{equation}
The attractive interaction of baryons with scalar field $\sigma$ modifies their masses as
\begin{equation}
M_{j}^{*}=M_{j}-g_{\sigma jj}\sigma,
\end{equation}
and the pressure is 
\begin{equation}
p_{\sigma}(\sigma)=-\frac{1}{2}m_{\sigma}^{2}\sigma_{0}^{2}.
\end{equation}
In order to determine the effective mass of the different species, we have to determine the mean scalar field $\sigma$. Using the thermodynamic consistency condition we can derive the following expression for the scalar field $\sigma$ as follows
\begin{equation}
\sigma=\frac{1}{m_{\sigma}^{2}}\sum_{j}g_{\sigma jj}(n_{\sigma j}+n_{\sigma\bar{j}}),
\end{equation}
where
\begin{equation}
n_{\sigma j(\bar{j})}=g_{j}\int\frac{d^{3}k}{(2\pi)^3}\frac{M_{j}^{*}}{(M_{j}^{*2}+k^{2})^{1/2}}f_{j(\bar{j})}
\end{equation}
It is obvious to see that the solution for $M^{*}$ involves a set of coupled and self consistent equations and hence we have to solve the following set of seven coupled equations self consistently in order to get the values of the baryon densities and effective masses of different species. These equations are
\begin{eqnarray}
M_{N}^{*}&=&M_{N}-\left(\frac{g_{\sigma NN}}{m_{\sigma}}\right)^{2}\Big[\frac{g_{N}}{(2\pi)^{3}}\int d^{3}k\frac{M_{j}^{*}}{(M_{j}^{*2}+k^{2})^{1/2}}(f_{N}+f_{\bar{N}})\nonumber
\\
&+&\frac{g_{\Delta}}{(2\pi)^{3}}\int_{0}^{\infty}dM W(M)\int d^{3}k\frac{M_{\Delta}^{*}}{(M_{\Delta}^{*2}+k^{2})^{1/2}}(f_{\Delta}+f_{\bar{\Delta}})\nonumber
\\
&+&\frac{2}{3}\Big(g_{\Lambda}\int\frac{d^{3}k}{(2\pi)^3}\frac{M_{\Lambda}^{*}}{(M_{\Lambda}^{*2}+k^{2})^{1/2}}(f_{\Lambda}+f_{\bar{\Lambda}})\\
&+&\displaystyle{g_{\sum}}\int\frac{d^{3}k}{(2\pi)^3}\frac{M_{\sum}^{*}}{(M_{\sum}^{*2}+k^{2})^{1/2}}(f_{\sum}+f_{\bar{\sum}})\Big)\nonumber
\\
&+&\frac{1}{3}{g_{\Xi}\int\frac{d^{3}k}{(2\pi)^3}\frac{M_{\Xi}^{*}}{(M_{\Xi}^{*2}+k^{2})^{1/2}}(f_{\Xi}+f_{\bar{\Xi}})}\Big],\nonumber
\end{eqnarray}
and five equations for baryon density $n_{BN}, n_{B\Lambda}, n_{B\sum}, n_{B\Xi}, n_{B\Delta}$ as given by Eq. (20) and (22), respectively and one more equation represented by Eq. (17). Further, we get the effective masses of hyperons and $\Delta$ particles as follows
\begin{eqnarray}
M_{\Lambda,\sum}^{*}&=& M_{\Lambda,\sum}-\frac{2}{3}(M_{N}-M_{N}^{*})\nonumber
\\
M_{\Xi}^{*}&=& M_{\Xi}-\frac{1}{3}(M_{N}-M_{N}^{*})
\\
M_{\Delta}^{*}&=& M_{\Delta}-(M_{N}-M_{N}^{*}).\nonumber
\end{eqnarray}
For meson $m$, we have used the following ideal gas equation for its number density:
\begin{equation}
n_{m(\bar{m})}=g_{m}\int\frac{d^{3}k}{(2\pi)^3}\big(exp\big[\frac{(M_{m}^{*2}+k^{2})^{1/2}\mp \mu_{m}}{T}\big]-1\big)^{-1}.
\end{equation}
In order to get the final result for Eq. (13), we impose the condition of strangeness neutrality to get the pressure of HG. It is obvious that we cannot include more baryons into the HG spectrum because the calculation becomes too much complicated to handle. 
\section{Quasiparticle model (QPM)}
We have used a thermodynamically consistent quasiparticle description as proposed by Bannur in order to study the EOS of QGP [31]. In this model, the system of interacting massless quarks and gluons can be effectively described as an ideal gas of ``massive'' noninteracting quasiparticles. The mass of these quasiparticles depends explicitly on $T$ and implicitly on $\mu_{q}$ via QCD running coupling constant. In this model, we start with the definition of average energy and average number of particles and derive all the thermodynamical quantities from them in a consistent manner. The effective mass of the gluon changes with T and $\mu_{q}$ as follows [32]:
\begin{equation}
m_{g}^{2}(T)=\frac{N_c}{6} g^{2}(T) T^{2} \left(1+\frac{N_{f}^{'}}{6}\right),
\end{equation}
\noindent
where $N_c$ represents the number of colours. We have taken $N_c=3$ in our calculation and:
\begin{equation}
N_{f}^{'}=N_{f}+\frac{3}{\pi^2}\sum_{f}\frac{\mu_f^2}{T^2}.
\end{equation}
\noindent
Here $N_f$ is the number of flavours of quarks and $\mu_f$ is the quark chemical potential belonging to the flavour f. Similarly the effective mass of the quarks involves the following relation [31]:
\begin{equation}
m_{q}^{2}=m_{q0}^{2}+\sqrt{2}m_{q0}m_{th}+m_{th}^{2},
\end{equation}
\noindent
were $m_{q0}$ is the rest mass of the quarks. In this calculation, we have used $m_{q0}=8 MeV$ for two light quarks ({\it u,d}), and $m_{q0}=80 MeV$ for strange quark. In the above Eq (34) $m_{th}$ represents the thermal mass of the quarks and it can be written as [33]:
\begin{equation}
m_{th}^{2}(T,\mu)=\frac{N_{c}^{2}-1}{8 N_{c}}\left[T^{2}+\frac{\mu_{q}^{2}}{\pi^2}\right]g^{2}(T),
\end{equation}
\noindent
Taking these values for the effective masses, energy density can be derived from the grand canonical  partition function in a thermodynamically consistent manner and is given as [34]:
\begin{equation}
\epsilon=\frac{T^4}{\pi^2}\sum_{l=1}^{\infty}\frac{1}{l^4}\left[\frac{d_g}{2}\epsilon_{g}(x_{g}l)+(-1)^{l-1}d_{q}cosh(\mu_{q}/T)\epsilon(x_{q}l)+(-1)^{l-1}\frac{d_{s}}{2}\epsilon_{s}(x_{s}l)\right],
\end{equation}
\noindent
with $\epsilon_{i}(x_{i}l)=(x_{i}l)^{3}K_{1}(x_{i}l)+3 (x_{i}l)^{2}K_{2}(x_{i}l)$, where $K_1$ and $K_2$ are the modified Bessel functions with $x_{i}=\frac{m_i}{T}$ and index i runs for gluons, up-down quarks q, and strange quark s. Here $d_i$ are the degeneracies associated with the internal degrees of freedom. Now, by using the thermodynamic relation $\epsilon=T\frac{\partial \it {p}}{\partial T}-\it {p}$, pressure of system at $\mu_{q}=0$ can be obtained as:
\begin{equation}
\frac{\it{p}(T,\mu_{q}=0)}{T}=\frac{\it{p}_0}{T_0}+\int_{T_0}^{T}dT \frac{\epsilon(T,\mu_{q}=0)}{T^2},
\end{equation}
\noindent
where $\it{p}_0$ is the pressure at a reference temperature $T_0$. We have used $\it{p}_{0}$=0 at $T_{0}$=100 MeV in our calculation. We get the pressure for a system at finite $\mu_{q}$ 
\begin{equation}
\it{p}(T,\mu_{q})=\it{p}(T,0)+\int_{0}^{\mu_{q}}n_{q}d\mu_{q}.
\end{equation}
\noindent
where the expression for $n_{q}$ is:
\begin{equation}
n_{q}=\frac{d_{q}T^{3}}{\pi^2}\sum_{l=1}^{\infty}(-1)^{l-1}\frac{1}{l^3}sinh(\mu_{q}/T)I_{i}(x_{i}l)
\end {equation}
\noindent
with $I_{i}(x_{i}l)=(x_{i}l)^2 K_{2}(x_{i}l)$. Thus all the thermodynamical quantities can be obtained in a consistent way by using this model. We have used the following expression for the coupling constant [18]
\begin{equation}
\alpha_{S}(T)=\frac{g^{2}(T)}{4 \pi}=\frac{6 \pi}{\left(33-2 N_{f}\right)\ln \left(\frac{T}{\Lambda_{T}}\sqrt{1+a\frac{\mu_{q}^{2}}{T^2}}\right)}
\\
\left(1-\frac{3\left(153-19 N_f \right)}{\left(33-2 N_f\right)^2}\frac{\ln \left(2 \ln \frac{T}{\Lambda_T}\sqrt{1+a\frac{\mu_{q}^{2}}{T^2}} \right)}{\ln \left(\frac{T}{\Lambda_{T}}\sqrt{1+a\frac{\mu_{q}^{2}}{T^2}}\right) }\right),
\end{equation}
where $\Lambda_{T}=115 MeV$ and $a=\frac{1}{\pi^{2}}$.
\begin{figure}
\includegraphics[height=28em]{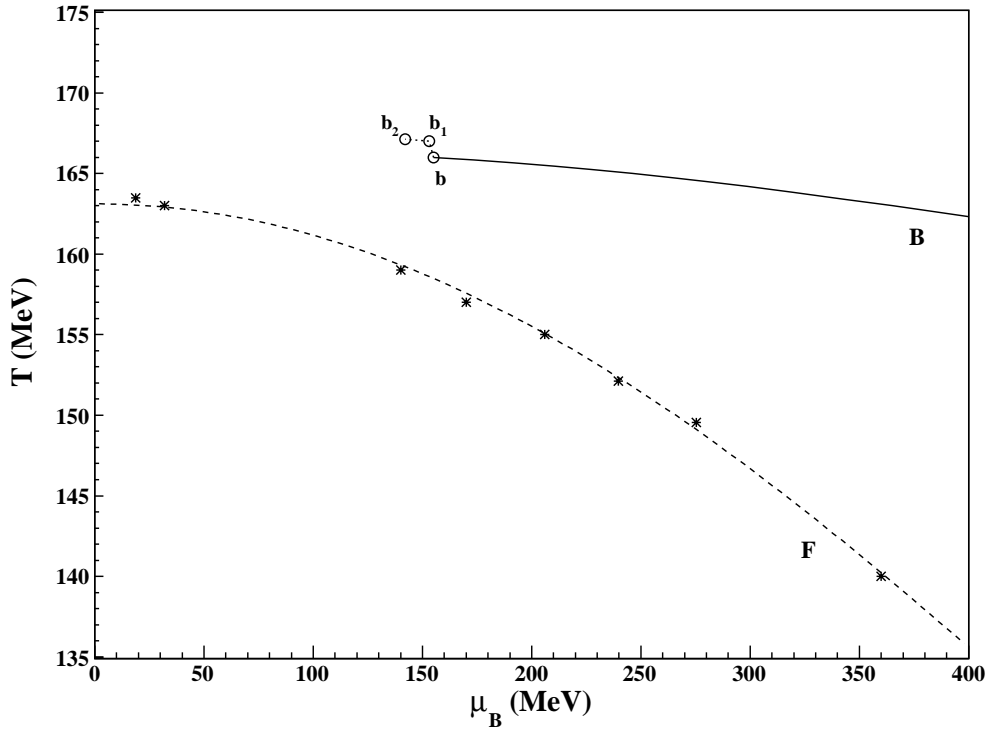}
\caption[]{QCD phase diagram in $T-\mu_{B}$ plane. F is the freezeout curve obtained from our excluded volume model for HG. B is the first order deconfinement phase transition line using EOS in the QPM and EOS for HG in our excluded volume model. Open points $b, b_{1}$ and $b_{2}$ are the locations of CEP for baryon's hard-core radius $r=0.8, r=0.6$ and $r=0.4 fm$, respectively.}
\end{figure}

\begin{figure}
\includegraphics[height=28em]{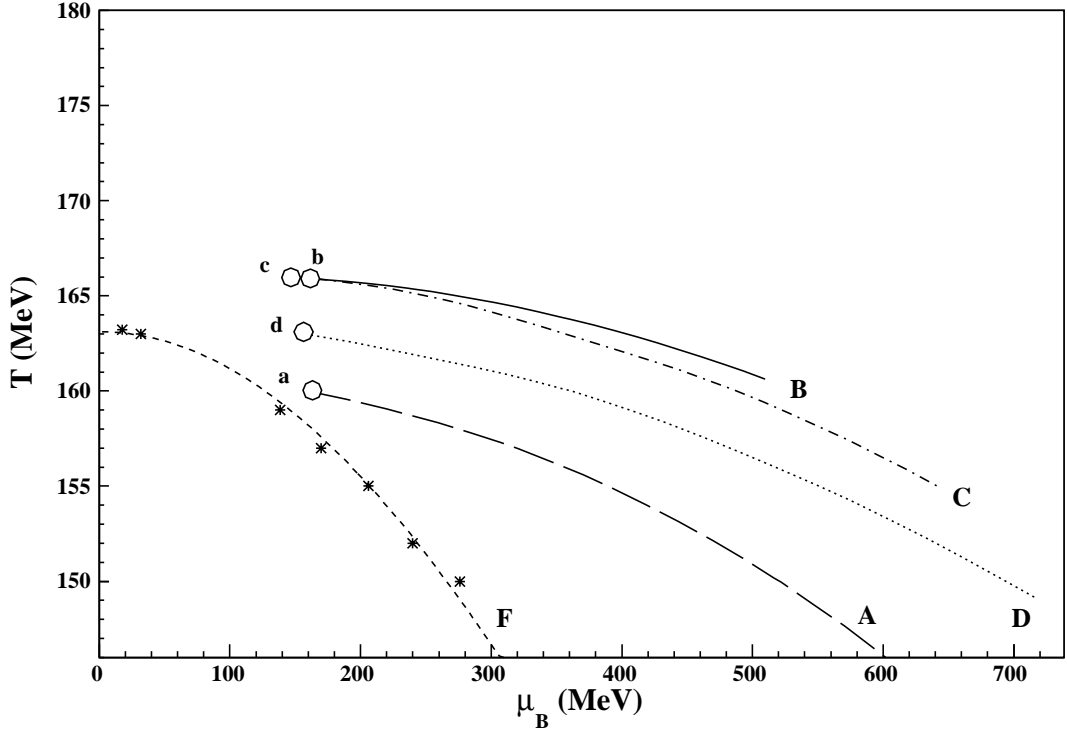}
\caption[]{QCD phase diagram in $T-\mu_{B}$ plane. F is the freezeout curve obtained from our excluded volume model for HG. A is the phase boundary using bag model for the EOS of QGP [17]and our EOS for HG. The a gives the location of CEP. B is the boundary using EOS in QPM and our excluded volume model for HG and b gives CEP. C is the first order deconfinement phase transition line using QPM and the simple Cleymans and Suhonen model for HG and c is the end point of this curve. Similarly, D is the line obtained using QPM and mean field model for HG and d is the corresponding end point.}
\end{figure}

\begin{figure}
 \includegraphics[height=20em]{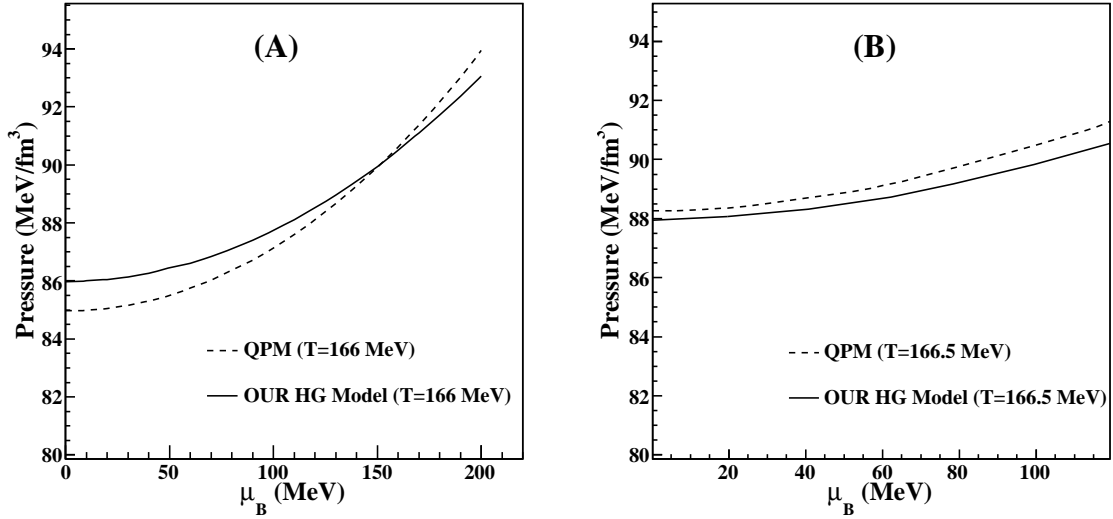}
\caption[]{Variation of pressure with respect to $\mu_{B}$ at different temperatures near CEP. Solid curve is the result for HG from our excluded volume model and dashed curve is the pressure for QGP obtained from QPM. Pressure equality condition is not fulfilled if $T$ is increased by $0.5 MeV$ from $T=166 MeV$.}
\end{figure}
\section{Results and Discussion}

\begin{figure}
\begin{center}
\noindent
\includegraphics[height=28em]{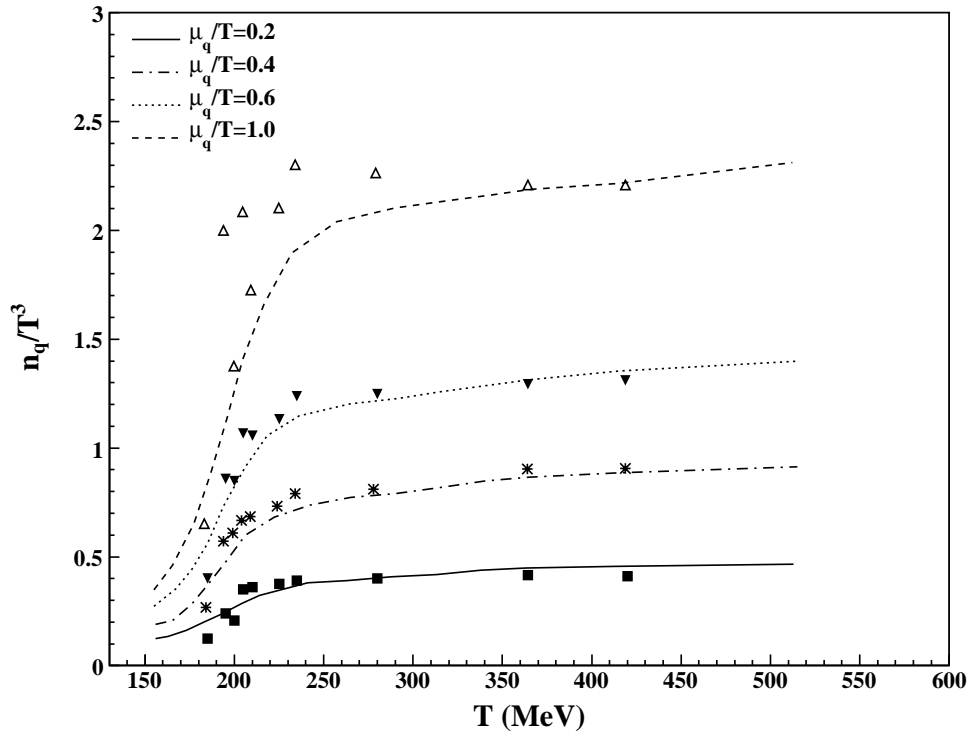}
\caption[]{Variation of normalized net quark density with respect to temperature at different $\mu_{q}/T$ [35]. Different points are the lattice data from Ref. [36].}
\end{center}
\end{figure}
\begin{table}
\begin{center}
\caption{Ratio of $n_{B}/n$ at the CEP using HG and quasiparticle model.}
\begin{tabular}{cccccc} \hline \hline 
HG Models             &coordinates of CEP   & $r$            & $n_{B}$               & $n$                  & $n_{B}/n$ \\ 
                      &$(T, \mu_{B})$       & $(fm)$         & $(fm^{-3})$           & $(fm^{-3})$          &    \\ \hline
Cleymans and Suhonen  & (166, 149)          & 0.8            & 0.163                 & 0.86                 &0.190\\
Our HG Model          & (166, 155)          & 0.8            & 0.104                 & 0.54                 &0.192\\
Mean field            & (163, 157)          & 0.8            & 0.0981                & 0.492                &0.199\\\hline
\end{tabular}
\end{center}
\end{table}

\begin{figure}
\includegraphics[height=28em]{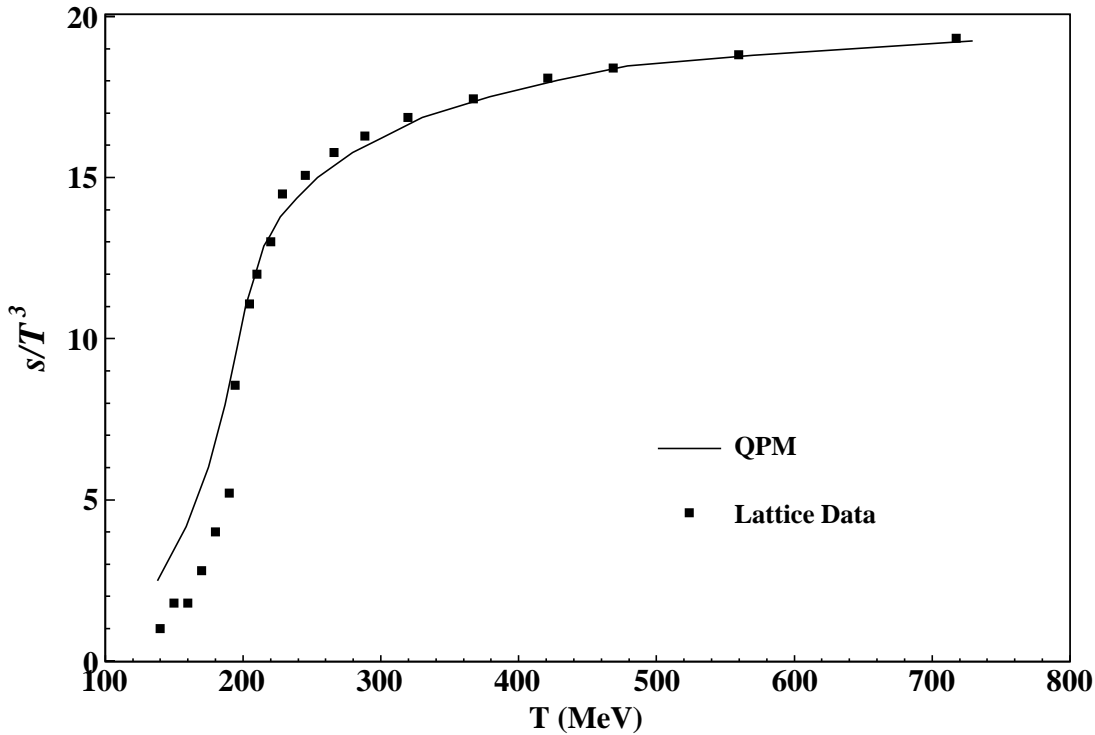}
\caption[]{Variation of normalized entropy density with respect to temperature at $\mu_B=0.0$ in QPM. Different points are the lattice data from Ref.[37].}
\end{figure}

\begin{figure}
\includegraphics[height=28em]{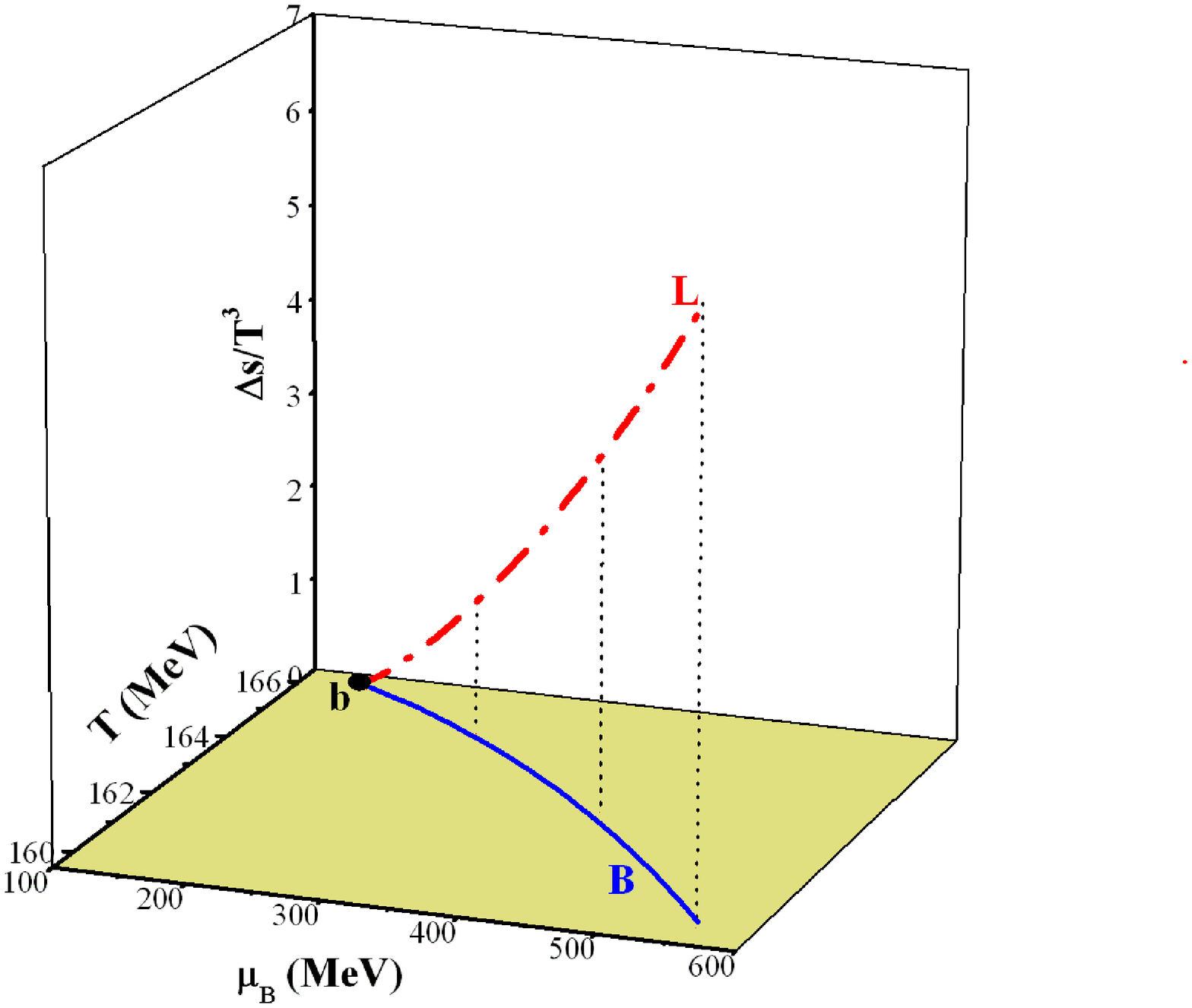}
\caption[]{Variation of $ (\Delta s/T^{3})=(s/T^{3})_{QGP}- (s/T^{3})_{HG}$ with respect to coordinates of various phase transition points on the $(T, \mu_{B})$ phase boundary. We have used transition points from the curve B of Fig. 2.}
\end{figure}
 In Fig. 1, we have demonstrated the location of the CEP when we adapt QPM as the EOS of the QCD plasma and also use our excluded volume model for the description of HG. The open points $b, b_{1}, b_{2}$ on the curve represent the variations in the coordinates of CEP when we change the hard-core radius of the baryons as $r=0.8 fm$, $r=0.6 fm$ and $r=0.4 fm$, respectively. Thus although CEP has its origin in the excluded volume effect, the drastic alterations in the hard-core radii do not yield much change in the location of CEP. Moreover, its proximity to the freezeout curve calculated in our HG model also remains unaltered.

In Fig. 2, we have shown the variations in the phase boundary when we either change the EOS of the QCD plasma or we vary the EOS of HG. If we use simplistic bag model for the EOS of plasma and our excluded volume model for HG, we get the phase boundary represented by the curve A and its terminal point $a$ is the location of CEP. It lies closest to the freezeout curve. Similarly B, C and D represent the phase boundaries when we use QPM for the EOS of QGP, but the EOS of HG is respectively taken in the form of our excluded volume model, Cleymans and Suhonen excluded volume model, and mean-field model with excluded volume correction. Their terminal points $b$, $c$, and $d$ represent the locations of CEP in these models, respectively. This exercise although results in a sizable variation in the coordinates of CEP, but the effect due to the details of the models is still found to be small.

In Fig. 3, the details of the pressure equality condition $P_{QGP}=P_{HG}$ at and around CEP are shown when we model QGP by QPM and HG by our excluded volume model. This clearly shows that this condition is very sensitive to a small variation of $0.5 MeV$ in the temperature when we find that the condition fails to hold and the deconfining transition does not occur. It also defines the beginning of a cross-over region lying beyond the critical end point where the meson dominant HG pressure is always less than the QGP pressure. Hence the dual description with quarks, gluons together with $\pi$ and $K$-mesons as constituents applies well in this region. The main assumption in our model is that the pressure of HG is reduced even if a large number of hadrons is produced at large T and/or $\mu_{B}$. It means that at large $\mu_{B}$, the fractional occupied volume R in our model increases and hence mobility of the baryons is considerably reduced. However, in our consideration mesons do not possess any such hard-core volume and hence they can fuse into one another when compressed. Thus we think of a possible parameter $x$ which defines the ratio of baryon density $n_{B}$ to the total number density $n (=n_{B}+n_{m})$ in the HG at the CEP. Here $n_{m}$ is the meson number density. In table 1, we show the values of this ratio at CEP obtained in various prescriptions of HG we have adopted here. Surprisingly we find that the ratio $x=0.195\pm 0.005$ which signifies that at CEP, the ratio $n_{m}:n_{B}$ is almost fixed as $4:1$ and is independent of HG models used in the calculations. Beyond CEP, the meson density increases and thus yields more dominant contribution in the cross-over region. This demonstrates that the location of CEP on the phase boundary requires that all the baryons in the HG possess a hard core volume but mesons when being compressed, can fuse into each other. Thus $n_{B}/n_{m}=0.25$ gives the location of CEP and this condition is independent of HG prescription used for the calculation. It still remains worth investigating problem why and how $n_{B}/n_{m}=0.25$ yields the precise location of CEP. In the excluded volume model, this ratio defines a critical fractional occupied volume $R_{C}$. As $\mu_{B}$ increases, the fractional occupied volume $R$ by baryons increases and consequently the mobility of baryons in the hot and dense HG decreases fast and it finally results into a reduced pressure of HG so that the Gibbs' conditions of equilibrium phase transition become satisfied. 

We must emphasize that we have used two different descriptions for QGP and HG, respectively. We find that the earlier version of our model [21] for the HG used with the Boltzmann approximation describes well the ratios of various particle multiplicities and we hope that the present version with full quantum statistics will still improve the results of comparison with the experimental data. In order to have confidence in our EOS for QGP, we must test its predictions with the recent lattice results obtained at zero as well as finite baryon density.  

In Fig. 4 and 5, we have shown the variations of normalized net quark density $n_{q}/T^{3}$ and normalized entropy density $s/T^{3}$ with respect to $T$ at different $\mu_{q}$ in the quasiparticle model (QPM). We find that our results yield a good fit to the lattice results. This comparison with the lattice calculation shows that QPM together with its parameters used here indeed gives a proper EOS for QGP even at finite $\mu_{B}$. In Fig. 6, we have attempted to show what happens to the change in the entropy density at CEP. We have calculated the difference $\frac{\Delta s}{T^{3}}=(s/T^{3})_{QGP}-(s/T^{3})_{HG}$ and demonstrated its variation with respect to the coordinates of the phase transition points lying at the boundary of Fig. 2. We find that $\frac{\Delta s}{T^{3}}=0.0$ at the CEP and is minimum. Although we have not yet established that CEP obtained in our calculation is a second-order phase transition point. Most importantly it is the terminal point of the phase boundary. However, our results clearly indicate that it can either give an isentropic or a second order phase transition point. This is certainly an interesting finding. Although we have used two different models for the description of QGP and HG phases, the vanishing of net entropy density at the CEP outlines a continuity in these descriptions.

The physical mechanism involved in this calculation is intuitively analogous to the percolation model where also a first order phase transition results with 'jamming' of baryons and thus mobility of baryons is affected [38-39]. However, in the percolation model we do not have any comparison to what we should get in the QGP picture. Here we use a similar picture and we explicitly and separately consider both the phases, i.e., HG as well as QGP and hence it gives a clear understanding how a first-order deconfining phase transition can be constructed in nature and finally we reach an interesting finding that the baryonic size is crucially responsible for the existence of CEP on the phase boundary in such a construction. At low baryon density, overlapping mesons fuse into each other and form a large bag or cluster, whereas at high baryon density, hard-core repulsion among baryons, restricts the mobility of baryons. Consequently we consider two distinct limiting regimes of HG, one beyond CEP is meson-dominant regime and the other is baryon dominant region.

A question generally arises : does our calculation offer any intuitive mechanism regarding the origin of the cross-over region? Beyond CEP, the cross-over region naturally appears in our model when $T$ further increases and $\mu_{B}$ decreases. Cross-over is defined as a gradual change of the system from one phase to the other without a definite transition point. Lattice QCD has confirmed the existence of the cross-over region at $\mu_{B}=0$ between HG and QGP. However, what happens in this region is still an open question so far as QCD is concerned. QCD involves two distinct vacua usually called as perturbative and physical one. Cross-over is thus realised by a gradual transition from one vacuum to the other. However, decomposition of hadrons to quarks and antiquarks one by one contradicts colour confinement because an isolated coloured object cannot exist in a physical vacuum [40]. In Nambu-Jona-Lasinio (NJL) as well as Polyakov extended Nambu-Jona-Lasinio (PNJL) models, the thermodynamic potential involves two degenerate minima at which two phases are in thermal, mechanical and chemical equilibria according to the Gibbs' criteria for the first order phase transition between the phases of broken and restored symmetry [41]. At the CEP temperature $T_{CEP}$, the chiral transition changes to the second order. For $T>T_{CEP}$, the thermodynamic potential has only one minimum and the transition is a smooth cross-over. However, the mechanism of cross-over is not understood very well because colour confinement does not strictly hold at the transition in these models. Our model falls in line with the ideas proposed recently [42-43] where it was shown that under circumstances, hot and dense HG consisting of extended hadrons could produce phase transition of the first or second order and also a smooth cross-over. We propose that although each baryon possesses a hard-core size, mesons are also extended particles but they lack a hard-core size. So they can overlap, fuse and interpenetrate. At CEP, mesons and baryons saturate the volume of the hot fireball. In meson dominated region (i.e., $T>T_{CEP}$), mesons have a far larger density than that of baryons. When they start overlapping on each other, they fuse into one another and cluster formation starts where colour can flow and only the cluster as a whole is colour-singlet. As the clusters merge together resulting into an infinitely sized cluster, analytic cross-over into a new phase occurs. Essentially we assume that each hadron is an extended bag of QGP and thus cluster formation arising due to fusion of mostly pions at $T>T_{CEP}$, creates a smooth cross-over transition from one phase to the other. This picture appears more clear when we consider HG at $\mu_{B}=0$. However, we must emphasize that unlike other effective models, we use a hybrid model where EOS for HG and QGP are constructed independently and they reproduce separately the experimental particle-multiplicity data as well as lattice QCD results, respectively.  It should be added here that many authors in the past have used two different equations of state for QGP and HG and obtained an explanation to an analytic and smooth cross-over and CEP in their models [44-45]. Our model presents a similar picture. Matching of the pressures at the CEP transition in the hybrid model used by us throws light on the mechanism of cross-over transition and as mentioned above, it is controlled by the presence of baryons in the system. But why does the ratio $n_{B}/n_{m}=0.25$ (a fixed value) at the CEP? How does the presence of baryon density affect the cluster formation? These questions need a thorough investigation before we make a clear picture.

In conclusion, searching for the precise location of the critical end point (CEP) in the QCD phase diagram still poses a challenging problem. Although various calculations have predicted its existence but the quantitative predictions regarding its location wildly differ. Experiments face an uphill task in probing the CEP in QCD phase diagram because a clarity in theoretical prediction is missing. Moreover, many unstudied problems such as short lifetime and the reduced volume of the QGP formed at colliders also affect the location of CEP and its verification [46]. In these circumstances, our results arising due to baryon size, will be helpful in understanding the origin of CEP and determining its location on the phase diagram.

\section {Acknowledgments}
 PKS and SKT are grateful to the University Grants Commission (UGC) and Council of Scientific and Industrial Research (CSIR), New Delhi for providing a research fellowship. CPS acknowledges the financial support through a project sanctioned by Department of Science and Technology, Government of India, New Delhi.

\pagebreak

\end{document}